# Refined Thermodynamic Uncertainty Relation for Chemical Reactions

Author names:

Ryohei Yuno and Katsuaki Tanabe[*]

Author address:

Department of Chemical Engineering, Kyoto University, Nishikyo, Kyoto 615-8510, Japan

Corresponding author:

[*]Email: tanabe@cheme.kyoto-u.ac.jp

**Abstract**

Thermodynamic uncertainty relations elucidate the intricate balance between the precision of current and the thermodynamic costs or dissipation, marking a recent and enthralling advancement at the confluence of statistical mechanics, thermodynamics, and information theory. In this study, we derive a time–energy uncertainty relation tailored for chemical reactions, expressed in terms of the Gibbs free energy and chemical potential. This inequality holds true irrespective of whether the total substance of chemical species is conserved during the reaction. Furthermore, it supports the general thermodynamic framework by ensuring the spontaneous decrease in Gibbs free energy. We present two formulations of the thermodynamic uncertainty relation: one based on chemical species concentrations and the other on molar fractions. The validity of our inequalities is numerically demonstrated using model systems of the Belousov–Zhabotinsky and Michaelis–Menten reactions. Our uncertainty relation may find practical applications in measuring and optimizing thermodynamic properties relevant to chemical reaction systems out of equilibrium.

# 1. Introduction

Nonequilibrium statistical mechanics and thermodynamics constitute a comprehensive framework for understanding the energetics and dynamics of stochastic processes occurring far from equilibrium. Fluctuations, which manifest as random deviations around the mean values of thermodynamic quantities, are pivotal in elucidating the dynamic behavior of such systems. The recent advancement of thermodynamic uncertainty relations, situated at the crossroads of statistical mechanics, thermodynamics, and information theory, delineates the trade-offs between current precision and thermodynamic cost or dissipation [1–8]. Beyond theoretical implications, the thermodynamic uncertainty relations find practical applications in the optimization of experimental setups and the development of protocols for the precise measurement of thermodynamic properties in nonequilibrium systems. Their potential applications span diverse nonequilibrium systems, such as heat engines [9], molecular motors [10], self-assembly phenomena [11], and anomalous diffusion [12]. By utilizing a thermodynamic uncertainty relation, one can, for instance, estimate the entropy production by measuring experimentally accessible currents and their fluctuations, obviating the need for knowledge about the interaction potentials or driving forces [13–15].

Based on the thermodynamic uncertainty relations presented by Nicholson *et al.* [16] and by Yoshimura and Ito [17], we previously derived a time–energy uncertainty relation in chemical thermodynamics in terms of the Gibbs free energy and chemical potential for constant pressure and temperature systems,

$$|\dot{G}| \leq \Delta\dot{\mu}\Delta\mu/RT , \quad (1)$$

which is eq. 26 of Ref. 18 and where *G* is the Gibbs free energy, *μ* is the chemical potential, *R* is the gas constant, and *T* is the temperature [18]. The dot atop a variable denotes the evolution rate or time derivative of the variable. Δ does the standard deviation of the variable. However, as we will describe later, this uncertainty relation was only applicable to chemical reactions in which the total amount of substance is conserved. In addition, due to the suboptimal setting of the standard chemical potential, *G* did not consistently adhere to the conventional thermodynamic description, failing to exhibit spontaneous decrease in all instances. To address these issues, in the present study, we refine our uncertainty relation to enhance its applicability across more general scenarios.

## 2. Chemical Reaction Models for Numerical Simulations

*2.1 Belousov–Zhabotinsky reaction*

In this section, we introduce the chemical reaction models we employ in this study to numerically test the thermodynamic uncertainty relations. Firstly, the Belousov–Zhabotinsky reaction consists of the following chemical reactions [17–21]:

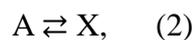   (2)

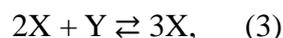   (3)

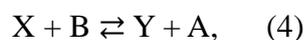   (4)

where A, B, X, and Y are chemical species. The reaction rates, $J_1$, $J_2$, and $J_3$ for Eqs. (2), (3), and (4), respectively, in the forward direction from the left- to right-hand side of each equation, are formulated as

$$J_1 = k_1^+[A] - k_1^-[X], \quad (5)$$

$$J_2 = k_2^+[X]^2[Y] - k_2^-[X]^3, \quad (6)$$

$$J_3 = k_3^+[X][B] - k_3^-[Y][A], \quad (7)$$

where [A], [B], [X], and [Y] are the concentrations of the chemical species, and $k_i^+$ and $k_i^-$ ($i = 1, 2, 3$) are the reaction rate constants in the forward and backward reactions, respectively, of the $i$th chemical equation. Then, the time derivatives of the chemical species' concentrations are:

$$[\dot{X}] = J_1 + J_2 - J_3, \quad (8)$$

$$[\dot{Y}] = -J_2 + J_3, \quad (9)$$

$$[\dot{A}] = -J_1 + J_3, \quad (10)$$

$$[\dot{B}] = -J_3. \quad (11)$$

In our calculations, the values of the reaction rate constants were set as: $k_1^+ = 1 \times 10^{-3}$, $k_1^- = 1$, $k_2^+ = 1$, $k_2^- = 1$, $k_3^+ = 1 \times 10^{-2}$, and $k_3^- = 1 \times 10^{-4}$. The initial values of the chemical species' concentrations were: $[A]_0 = 1 \times 10^3$, $[B]_0 = 1 \times 10^3$, $[X]_0 = 1$, and $[Y]_0 = 6$ [17,18].

Importantly, in this whole reaction system, the total amount of substance is conserved throughout the reactions, as the amount of substance of the product(s) is equal to that of the reactant(s) in each of Eqs. (2) – (4). Figure 1 plots the calculated time evolution of the concentrations [A], [B], [X], and [Y] of the chemical species A, B, X, and Y, respectively. It is observed that the concentrations [X] and [Y] are dynamically oscillating, and thus indicating the suitability of this reaction system for discussions in the field of nonequilibrium thermodynamics.

*2.2 Michaelis–Menten reaction*

Secondly, the Michaelis–Menten reaction is widely used in biochemistry to describe enzymatic reactions in solution. The involved reactions are [22,23]:

$$E + S \rightleftarrows ES, \quad (12)$$

$$ES \rightleftarrows E + P, \quad (13)$$

where E is the enzyme, S is the substrate, ES is the enzyme–substrate complex, and P is the product. The reaction rates, $J_1$, and $J_2$ for Eqs. (12) and (13), respectively, in the forward direction from the left- to right-hand side of each equation, are formulated as

$$J_1 = k_1^+[E][S] - k_1^-[ES], \quad (14)$$

$$J_2 = k_2^+[ES] - k_2^-[E][P], \quad (15)$$

where [E], [S], [ES], and [P] are the concentrations of the enzyme, substrate, enzyme–substrate complex, and product, respectively. The time derivatives of the chemical species' concentrations are:

$$[\dot{E}] = -J_1 + J_2, \quad (16)$$

$$[\dot{S}] = -J_1, \quad (17)$$

$$[\dot{ES}] = J_1 - J_2, \quad (18)$$

$$[\dot{P}] = J_2. \quad (19)$$

We set the values of the reaction rate constants as: $k_1^+ = 1$, $k_1^- = 1$, $k_2^+ = 1$, and $k_2^- = 1 \times 10^{-4}$. The initial values of the chemical species' concentrations were set as: $[E]_0 = 10$, $[S]_0 = 10$, $[ES]_0 = 0$, and $[P]_0 = 0$. It should be note that the total amount of substance is not conserved during the reaction. Figure 2 plots the calculated time evolution of the concentrations [E], [S], [ES], and [P].

## 3. Theory, Formulation, and Validation

*3.1 Issues in the previous formulation*

In Ref. 18, we derived a thermodynamic uncertainty relation for chemical reactions, described as Eq. (1) above. The constituents in our previous formulation were:

$$\dot{G} = RT \sum \dot{p}_i \ln p_i, \quad (20)$$

$$\Delta\dot{\mu} = \sqrt{\sum p_i \left(\dot{\mu}_i - \langle\dot{\mu}\rangle\right)^2} = \sqrt{\sum p_i \left(\dot{\mu}_i - \sum p_i \dot{\mu}_i\right)^2}, \quad (21)$$

and

$$\Delta\mu = \sqrt{\sum p_i \left(\mu_i - \langle\mu\rangle\right)^2} = \sqrt{\sum p_i \left(\mu_i - \sum p_i \mu_i\right)^2}, \quad (22)$$

which correspond to eqs. 12, 24, and 16 of Ref. 18, respectively. $p_i$ is the molar fraction of the chemical species $X_i$ ($i = 1, 2, \ldots, N$), where $N$ is the number of chemical species involved in the chemical reaction system. It turns out that Eq. (20) was problematic, whose derivation is as follows. The time derivative of the Gibbs free energy for constant pressure and temperature processes is given as

$$\dot{G} = \sum [\dot{X}_i]\mu_i, \quad (23)$$

which is eq. 11 of Ref. 18 and where $[X_i]$ is the concentration of the chemical species $X_i$. The chemical potential of the chemical species $X_i$ is

$$\mu_i = \mu_i^0 + RT \ln[X_i] \quad (24)$$

for an ideal solution, which is eq. 9 of Ref. 18 and where $\mu_i^0$ is the standard chemical potential of $X_i$. Then, we derived Eq. (20) by simply substituting $[X_i]$ with $p_i$ and setting $\mu_i^0$ to zero. Under these definitions, the thermodynamic uncertainty relation of Eq. (1) holds for chemical reactions that conserve the total amount of substance (i.e., the amount of substance of the products equaling that of the reactants), as numerically demonstrated in Ref. 18. Nevertheless, in view of the consistency with the general description of thermodynamics and applicability, there were two issues in this formulation. Firstly, the Gibbs free energy $G$ does not always decrease spontaneously. Figure 3 presents the calculated time evolution of $\dot{G}$ and $G$ for the Belousov–Zhabotinsky reaction model. $R$ and $T$ were set to unity, and the initial value of $G$ was set to zero. $\dot{G}$ is occasionally observed to be positive values, and $G$ can consequently increases, which is inconsistent with the framework of general thermodynamics. This discrepancy stems from setting $\mu_i^0$ to zero, neglecting the requisite conditions for the standsard chemical potential, an aspect addressed in the subsequent subsection. Secondly, the thermodynamic uncertainty relation of Eq. (1) in this formulation is valid only for chemical reactions in which the total amount of substance is conserved, but not for non-conserved systems. This limitation arises from our simple replacement of $[X_i]$ with $p_i$. The concentration is the amount of substance of the chemical species divided by the total volume of the system. Whilst, the molar fraction is the amount of substance of the chemical species divided by the total amount of substance or the sum of the amount of substance of all chemical species in the system. The total volume can be regarded as a time-independent constant for dilute solutions. Therefore, while this substitution of the concentration with the molar fraction is valid for chemical reaction systems that conserve the amount of substance due to the volume's constancy, it's inadequate for non-conserved systems where the total amount of substance temporally changes. In order to solve this problem, we derive thermodynamic uncertainty relations that can be applied to chemical reactions with non-conserved total amount of substance in the subsequent subsections.

*3.2 Modification of the standard chemical potential*

First of all, we carefully reset the standard chemical potential $\mu_i^0$. The fundamental thermodynamic equation in terms of the Gibbs free energy goes

$$dG = VdP - SdT + \sum \mu_i dn_i, \quad (25)$$

where $V$, $P$, and $S$ are the total volume, pressure, and entropy of the system, respectively, and $n_i$ is the amount of substance of the chemical species $X_i$. Therefore, under the constant pressure and temperature condition,

$$dG = \sum \mu_i dn_i. \quad (26)$$

The infinitesimal evolution of the amount of substance of each chemical species is written in terms of the extent of reaction as

$$dn_i = \nu_i d\xi, \quad (27)$$

where $\nu_i$ is the stoichiometric number of the chemical species $X_i$ in the chemical reaction equation and $\xi$ is the extent of reaction. In the framework of general thermodynamics, the Gibbs free energy becomes minimum at equilibrium, and therefore,

$$\frac{dG}{d\xi} = 0. \quad (28)$$

From Eqs. (26) – (28),

$$\sum v_i \mu_i^{eq} = 0, \quad (29)$$

where $\mu_i^{eq}$ is the chemical potential of the chemical species $X_i$ at equilibrium, has to be satisfied for each reaction, as a prerequisite for the setting of $\mu_i^0$. Note that $n_i$ in Eqs. (25) and (26) could be replaced with $[X_i]$, which however does not eventually affect the conclusion of Eq. (29). Also note that Eq. (23) originally stems from Eq. (26). Under these $M$ constraints, where $M$ is the number of reactions, we still have a degree of freedom of $N-M$ for the choice of $\mu_i^0$. For example, one could set $\mu_i^0$ as zero for arbitrary $N-M$ chemical species and as specific values that satisfy Eq. (29) for the other $M$ chemical species. In the present study, we set $\mu_i^0$ so that all of $\mu_i$'s become zero at equilibrium for convenience. In this setting,

$$\mu_i^{eq} = \mu_i^0 + RT \ln[X_i]_{eq} = 0, \quad (30)$$

where $[X_i]_{eq}$ is the concentration of the chemical species $X_i$ at equilibrium. Therefore,

$$\mu_i^0 = -RT \ln[X_i]_{eq}. \quad (31)$$

Note that this setting of the standard chemical potential is not a necessary condition but a sufficient condition with a redundant degree of freedom of $N-M$. Similarly, for the description of chemical potential using the molar fraction of the chemical species,

$$\mu_i^0 = -RT \ln p_i^{eq}, \quad (32)$$

where $p_i^{eq}$ is the molar fraction of the chemical species $X_i$ at equilibrium.

*3.3 Concentration-based formulation*

Here we derive a thermodynamic uncertainty relation in terms of the Gibbs free energy and chemical potential. For the description of chemical potential using the concentration of the chemical species, Eq. (24), as Refs. 17 and 18,

$$\mu_i = RT \ln \frac{[X_i]}{[X_i]_{eq}} \quad (33)$$

by substituting Eq. (31) into Eq. (24). By substituting Eq. (33) into Eq. (23),

$$\dot{G} = RT \sum [\dot{X}_i] \ln \frac{[X_i]}{[X_i]_{eq}}. \quad (34)$$

Figure 4 plots the time evolution of $\dot{G}$ and $G$ computed using Eq. (34) for the Belousov–Zhabotinsky reaction model. $R$ and $T$ were set to unity, and the initial value of $G$ was set to zero. Unlike the case of Fig. 3, $\dot{G}$ is always observed to be negative and thus $G$ consistently decreases. This behavior aligns seamlessly with the framework of general thermodynamics.

Instead of the standard deviation of chemical potential based on the mean value of chemical potential employed in Ref. 18 as Eq. (22), we in the present study define the chemical standard deviation of chemical potential based on the equilibrium chemical potential as

$$\Delta\mu = \sqrt{\sum p_i \left(\mu_i - \mu_i^{eq}\right)^2}, \quad (35)$$

similar to the chemical variance of chemical potential in Ref. 17. Nevertheless, it should be noted that Eq. (35) is slightly different from the chemical variance of chemical potential defined in Ref. 17, where $[X_i]$ was employed instead of $p_i$. We will get back to this topic later in the section *4. Discussion*. Then, the chemical standard deviation of the time derivative of chemical potential is

$$\Delta\dot{\mu} = \sqrt{\sum p_i \left(\dot{\mu}_i - \dot{\mu}_i^{eq}\right)^2}. \quad (36)$$

By substituting Eqs. (30) and (33) into Eq. (35),

$$\Delta\mu = RT \sqrt{\sum p_i \left(\ln \frac{[X_i]}{[X_i]_{eq}}\right)^2}. \quad (37)$$

Since $p_i = [X_i]/\Sigma[X_i]$, Eq. (37) could be rewritten as

$$\Delta\mu = RT \sqrt{\frac{1}{\sum[X_i]} \sum [X_i]\left(\ln \frac{[X_i]}{[X_i]_{eq}}\right)^2}. \quad (38)$$

From Eq. (33),

$$\dot{\mu}_i = RT \frac{[\dot{X}_i]}{[X_i]}, \quad (39)$$

and therefore, by substituting this and Eq. (30) into Eq. (36),

$$\Delta \dot\mu = RT\sqrt{\sum p_i \left(\frac{[\dot X_i]}{[X_i]}\right)^2}. \quad (40)$$

Since $p_i = [X_i]/\Sigma[X_i]$, Eq. (40) can be rewritten as

$$\Delta \dot\mu = RT\sqrt{\frac{1}{\sum[X_i]}\sum \frac{[\dot X_i]^2}{[X_i]}}. \quad (41)$$

From Eqs. (34), (38), and (41),

$$\begin{aligned}
|\dot G| &= RT \left|\sum[\dot X_i]\ln\frac{[X_i]}{[X_i]_{eq}}\right| \\
&= RT\left(\sum[X_i]\right)\left|\sum\left(\frac{1}{\sqrt{\sum[X_i]}}\frac{[\dot X_i]}{\sqrt{[X_i]}}\right)\left(\frac{\sqrt{[X_i]}}{\sqrt{\sum[X_i]}}\ln\frac{[X_i]}{[X_i]_{eq}}\right)\right| \\
&\leq RT\left(\sum[X_i]\right)\sqrt{\sum\left(\frac{1}{\sqrt{\sum[X_i]}}\frac{[\dot X_i]}{\sqrt{[X_i]}}\right)^2 \sum\left(\frac{\sqrt{[X_i]}}{\sqrt{\sum[X_i]}}\ln\frac{[X_i]}{[X_i]_{eq}}\right)^2} \quad (42)\\
&= RT\left(\sum[X_i]\right)\sqrt{\frac{1}{\sum[X_i]}\sum\frac{[\dot X_i]^2}{[X_i]}}\sqrt{\frac{1}{\sum[X_i]}\sum[X_i]\left(\ln\frac{[X_i]}{[X_i]_{eq}}\right)^2} \\
&= \frac{\sum[X_i]}{RT}\Delta\dot\mu\Delta\mu
\end{aligned}$$

At the point of inequality, the Cauchy–Schwarz inequality was applied [24]. Thus,

$$|\dot G|\Big/\sum[X_i] \leq \Delta\dot\mu\Delta\mu/RT. \quad (43)$$

Figures 5 and 6 present the time evolution of $|\dot G|/\Sigma[X_i]$ and $\Delta\dot\mu\Delta\mu/RT$ calculated for the Belousov–Zhabotinsky and Michaelis–Menten reaction models, respectively, with $R$ and $T$ set to unity. These results provide numerical evidence supporting the validity of the inequality expressed in Eq. (43) for both types of chemical reaction systems—those conserving and those not conserving the total amount of substance. It should be noted that the inequality presented in Ref. 17 was also valid for both of the conserved and non-conserved cases.

*3.4 Molar fraction-based formulation*

Let us also derive the thermodynamic uncertainty relation for the description of chemical potential using the molar fraction of the chemical species. The description of chemical potential using the molar fraction of the chemical species is

$$\mu_i = \mu_i^0 + RT \ln p_i. \quad (44)$$

This description of chemical potential may be more common rather than the concentration-based one, Eq. (24), as a component in general thermodynamics in various science and engineering fields. From Eqs. (32) and (44),

$$\mu_i = RT \ln \frac{p_i}{p_i^{eq}}, \quad (45)$$

and therefore,

$$\dot{\mu}_i = RT \frac{\dot{p}_i}{p_i}. \quad (46)$$

From Eq. (26),

$$\dot{G} = \sum \mu_i \dot{n}_i, \quad (47)$$

and therefore, from Eq. (45),

$$\dot{G} = RT \sum \dot{n}_i \ln \frac{p_i}{p_i^{eq}}. \quad (48)$$

The time evolution of $\dot{G}$ and $G$ calculated by Eq. (48) for the Belousov–Zhabotinsky reaction model was necessarily equivalent to that presented in Fig. 4, where $\dot{G}$ always exhibited negativity and $G$ consistently decreased over time, aligning with the framework of general thermodynamics.

By substituting Eqs. (30) and (45) into Eq. (35),

$$\Delta \mu = RT \sqrt{\sum p_i \left( \ln \frac{p_i}{p_i^{eq}} \right)^2}. \quad (49)$$

By substituting Eqs. (30) and (46) into Eq. (36),

$$\Delta \dot{\mu} = RT \sqrt{\sum \frac{\dot{p}_i^2}{p_i}}. \quad (50)$$

Accounting for the contents in Eqs. (48) – (50), $\dot{n}_i$ has to be converted to $\dot{p}_i$ in a straightforward way, to construct an uncertainty relation. Therefore, here we introduce the amount of substance of the solvent, $n_{sol}$. Since $p_i \cong n_i/n_{sol}$ for dilute solutions, $\dot{n}_i \cong n_{sol}\dot{p}_i$. Then, from Eqs. (48) – (50),

$$\begin{aligned}
|\dot{G}| &= RT\left|\sum \dot{n}_i \ln \frac{p_i}{p_i^{eq}}\right| \cong n_{sol}RT\left|\sum \dot{p}_i \ln \frac{p_i}{p_i^{eq}}\right| \\
&= n_{sol}RT\left|\sum \left(\frac{\dot{p}_i}{\sqrt{p_i}}\right)\left(\sqrt{p_i}\ln \frac{p_i}{p_i^{eq}}\right)\right| \\
&\leq n_{sol}RT\sqrt{\sum \left(\frac{\dot{p}_i}{\sqrt{p_i}}\right)^2 \sum \left(\sqrt{p_i}\ln \frac{p_i}{p_i^{eq}}\right)^2} \quad . \quad (51) \\
&= n_{sol}RT\sqrt{\sum \frac{\dot{p}_i^{\,2}}{p_i}}\sqrt{\sum p_i\left(\ln \frac{p_i}{p_i^{eq}}\right)^2} \\
&= \frac{n_{sol}}{RT}\Delta\dot{\mu}\Delta\mu
\end{aligned}$$

Thus,

$$|\dot{G}|/n_{sol} \leq \Delta\dot{\mu}\Delta\mu/RT. \quad (52)$$

Figures 7 and 8 present the computed time evolution of $|\dot{G}|/n_{sol}$ and $\Delta\dot{\mu}\Delta\mu/RT$ for the Belousov–Zhabotinsky and Michaelis–Menten reaction models, respectively. $n_{sol}$, $R$, and $T$ were set to unity. These results numerically demonstrated that the inequality of Eq. (52) holds for both of the cases of chemical reaction systems in which the total amount of substance is conserved and not conserved.

**4. Discussion**

To summarize, the main result of this study is the thermodynamic uncertainty relations of Eq. (43) (Eq. (52)) based on the concentration (molar fraction) of the chemical species, with the time derivative of Gibbs free energy of Eq. (34) (Eq. (48)), the chemical standard deviation of chemical potential of Eq. (38) (Eq. (49)), and the chemical standard deviation of the time derivative of chemical potential of Eq. (41) (Eq. (50)). To discuss the difference between the concentration-based (Eq. (43)) and molar fraction-based (Eq. (52)) formulations, it is important to note the difference in the unit of $G$, which is energy per volume in Eq. (43) and simply energy in Eq. (52). While the molar fraction-based description of chemical potential of Eq. (44) seems common as a component in general thermodynamics in various science and engineering fields, the concentration-based one of Eq. (24) is also popularly used in some specific communities such as electrochemistry. Therefore, one can employ a more suitable one out of the two types of formulations for convenience in each system under consideration.

As a practical significance of the thermodynamic uncertainty relations, for instance, the determination of the right-hand side values in Eqs. (43) and (52) can be relatively straightforward. This can be achieved, for example, by measuring the electric potential or voltage of the chemical solution along with its time derivative or evolution rate. Such measurements enable an estimation of the upper bound for the challenging Gibbs free energy rate. In this manner, the practical implications of this inequality may extend beyond theoretical constructs, offering tangible applications in optimizing experimental setups and devising protocols for the accurate assessment of thermodynamic properties in systems operating out of equilibrium. Such thermodynamic uncertainty relations could prove beneficial across diverse domains, including but not limited to heat engines, molecular motors, self-assembly phenomena, and anomalous diffusion [9–12]. Leveraging these relations allows one, for example, to gauge entropy production through the observation of experimentally measurable currents and their fluctuations, thereby circumventing the necessity for precise knowledge of interaction potentials or driving forces [13–15].

Here, instead of Eqs. (35) and (36), let us employ another form of standard deviation of chemical potential:

$$\Delta\mu = \sqrt{\sum [X_i](\mu_i - \mu_i^{eq})^2}, \quad (53)$$

which is based on the chemical variance of chemical potential defined in Ref. 17. Then, the chemical standard deviation of the time derivative of chemical potential is

$$\Delta\dot{\mu} = \sqrt{\sum [X_i](\dot{\mu}_i - \dot{\mu}_i^{eq})^2}. \quad (54)$$

By substituting Eqs. (30) and (33) into Eq. (53),

$$\Delta\mu = RT\sqrt{\sum [X_i]\left(\ln\frac{[X_i]}{[X_i]_{eq}}\right)^2}. \quad (55)$$

By substituting Eqs. (30) and (39) into Eq. (54),

$$\Delta\dot{\mu} = RT\sqrt{\sum \frac{[\dot{X}_i]^2}{[X_i]}}. \quad (56)$$

From Eqs. (34), (55), and (56),

$$\begin{aligned}
|\dot{G}| &= RT\left|\sum[\dot{X}_i]\ln\frac{[X_i]}{[X_i]_{eq}}\right| \\
&= RT\left|\sum\left(\frac{[\dot{X}_i]}{\sqrt{[X_i]}}\right)\left(\sqrt{[X_i]}\ln\frac{[X_i]}{[X_i]_{eq}}\right)\right| \\
&\leq RT\sqrt{\sum\left(\frac{[\dot{X}_i]}{\sqrt{[X_i]}}\right)^2 \sum\left(\sqrt{[X_i]}\ln\frac{[X_i]}{[X_i]_{eq}}\right)^2} \quad (57)\\
&= RT\sqrt{\sum\frac{[\dot{X}_i]^2}{[X_i]}}\sqrt{\sum[X_i]\left(\ln\frac{[X_i]}{[X_i]_{eq}}\right)^2} \\
&= \frac{\Delta\dot{\mu}\Delta\mu}{RT}
\end{aligned}$$

Thus,

$$|\dot{G}| \leq \Delta\dot{\mu}\Delta\mu/RT. \quad (58)$$

This thermodynamic uncertainty relation is eventually in the same form as Eq. (1), which is eq. 26 of Ref. 18, but with different and more appropriate constitutions of $\dot{G}$, $\Delta\dot{\mu}$, and $\Delta\mu$. Eq. (58) is also mathematically equivalent to eq. 79 of Ref. 17, where the Fisher information was employed instead of the standard deviation of the time derivative of chemical potential. The inequility of Eq. (58) is simpler than Eqs. (43) and (52), which represent Subsections 3.3 and 3.4, respectively. Nevertheless, $\Delta\mu$ in Eq. (53) has an awkward unit of $J\cdot mol^{-1/2}\cdot m^{-3/2}$ in the SI unit system, while Eq. (35) has is in $J\cdot mol^{-1}$, which is common as chemical potential in the field of engineering. Thus, one had better choose the most suitable thermodynamic unicertainty relation in each practical situation out of the three presented in this study, Eqs. (43), (52), and (58), accompanied with each definition of the standard deviation of chamical potential. Table 1 summarizes the bases of the chemical potential and standard deviation in the thermodynamic uncertainty relations presented in the present study, Refs. 17 and 18.

## 5. Conclusions

In this study, we derived formulations for the thermodynamic uncertainty relation, $|\dot{G}|/\sum[X_i] \leq \Delta\dot{\mu}\Delta\mu/RT$ (concentration-based) or $|\dot{G}|/n_{sol} \leq \Delta\dot{\mu}\Delta\mu/RT$ (molar fraction-based) for $\Delta\mu = \sqrt{\sum p_i (\mu_i - \mu_i^{eq})^2}$, and $|\dot{G}| \leq \Delta\dot{\mu}\Delta\mu/RT$ for $\Delta\mu = \sqrt{\sum [X_i](\mu_i - \mu_i^{eq})^2}$, under constant pressure and temperature conditions. These inequalities hold true whether the total amount of substance of the chemical species is conserved during the reaction or not, and also supports the general thermodynamic framework with the Gibbs free energy spontaneously decreasing. To validate the robustness of our inequalities, we conducted numerical analyses using model systems representative of the Belousov–Zhabotinsky and Michaelis–Menten reactions. The practical significance of our uncertainty relation is underscored by its potential applications in the measurement and optimization of thermodynamic properties associated with chemical reaction systems operating out of equilibrium.

**References**

[1] J. M. Horowitz and T. R. Gingrich, Nat. Phys. **16**, 15 (2020).

https://doi.org/10.1038/s41567-019-0702-6

[2] G. Falasco, M. Esposito, and J. C. Delvenne, New J. Phys. **22**, 053046 (2020).

https://doi.org/10.1088/1367-2630/ab8679

[3] T. Koyuk and U. Seifert, Phys. Rev. Lett. **129**, 210603 (2022).

https://doi.org/10.1103/PhysRevLett.129.210603

[4] N. J. López-Alamilla and R. U. L. Cachi, Chaos **32**, 103109 (2022).

https://doi.org/10.1063/5.0107554

[5] L. Ziyin and M. Ueda, Phys. Rev. Res. **5**, 013039 (2023).

https://doi.org/10.1103/PhysRevResearch.5.013039

[6] T. Van Vu and K. Saito, Phys. Rev. X **13**, 011013 (2023).

https://doi.org/10.1103/PhysRevX.13.011013

[7] T. Kamijima, S. Ito, A. Dechant, and T. Sagawa, Phys. Rev. E **107**, L052101 (2023).

https://doi.org/10.1103/PhysRevE.107.L052101

[8] Y. Hasegawa, Nat. Commun. **14**, 2828 (2023).

https://doi.org/10.1038/s41467-023-38074-8

[9] N. Shiraishi, K. Saito, and H. Tasaki, Phys. Rev. Lett. **117**, 190601 (2016).

https://doi.org/10.1103/PhysRevLett.117.190601

[10] P. Pietzonka, A. C. Barato, and U. Seifert, J. Stat. Mech., 124004 (2016).

https://doi.org/10.1088/1742-5468/2016/12/124004

[11] M. Nguyen and S. Vaikuntanathan, Proc. Natl. Acad. Sci. U.S.A. **113**, 14231 (2016).

https://doi.org/10.1073/pnas.1609983113

[12] D. Hartich and A. Godec, Phys. Rev. Lett. **127**, 080601 (2021).

https://doi.org/10.1103/PhysRevLett.127.080601

[13] J. Li, J. M. Horowitz, T. R. Gingrich, and N. Fakhri, Nat. Commun. **10**, 1666 (2019).

https://doi.org/10.1038/s41467-019-09631-x

[14] S. K. Manikandan, D. Gupta, and S. Krishnamurthy, Phys. Rev. Lett. **124**, 120603 (2020).

https://doi.org/10.1103/PhysRevLett.124.120603

[15] D. J. Skinner and J. Dunkel, Proc. Natl. Acad. Sci. U.S.A. **118**, e2024300118 (2021).

https://doi.org/10.1073/pnas.2024300118

[16] S. B. Nicholson, L. P. García-Pintos, A. del Campo, and J. R. Green, Nat. Phys. **16**, 1211 (2020).

https://doi.org/10.1038/s41567-020-0981-y

[17] K. Yoshimura and S. Ito, Phys. Rev. Res. **3**, 013175 (2021).

https://doi.org/10.1103/PhysRevResearch.3.013175

[18] K. Tanabe, AIP Adv. **12**, 035224 (2022).

https://doi.org/10.1063/5.0084251

[19] I. Prigogine and R. Lefever, J. Chem. Phys. **48**, 1695 (1968).

https://doi.org/10.1063/1.1668896

[20] J. J. Taboada, A. P. Munuzuri, V. Perez-Munuzuri, M. Gomez-Gesteira, and V. Perez-Villar, Chaos **4**, 519 (1994).

https://doi.org/10.1063/1.166030

[21] C. T. Hamik, N. Manz, and O. Steinbock, J. Phys. Chem. A **105**, 6144 (2001).

https://doi.org/10.1021/jp010270j

[22] A. R. Tzafriri, Bull. Math. Biol. **65**, 1111 (2003).

https://doi.org/10.1016/S0092-8240(03)00059-4

[23] K. Sekimoto, Stochastic Energetics, Springer, Heidelberg (2010).

https://doi.org/10.1007/978-3-642-05411-2

[24] K. Tanabe, IOP SciNotes **2**, 015202 (2021).

https://doi.org/10.1088/2633-1357/abe99f

**Figure and Table Captions**

**Figure 1:** (Color online) Time evolution of [A], [B], [X], and [Y] calculated for the Belousov–Zhabotinsky reaction model.

**Figure 2:** (Color online) Time evolution of [E], [S], [ES], and [P] calculated for the Michaelis–Menten reaction model.

**Figure 3:** (Color online) Time evolution of $\dot{G}$ and $G$ calculated based on the formulation in Ref. 18 for the Belousov–Zhabotinsky reaction model. The initial value of $G$ was set to zero.

**Figure 4:** (Color online) Time evolution of $\dot{G}$ and $G$ calculated based on Eq. (34) for the Belousov–Zhabotinsky reaction model. The initial value of $G$ was set to zero.

**Figure 5:** (Color online) Time evolution of $|\dot{G}|/\Sigma[X_i]$ and $\Delta\mu\Delta\dot{\mu}/RT$ calculated for the Belousov–Zhabotinsky reaction model, to demonstrate Eq. (43).

**Figure 6:** (Color online) Time evolution of $|\dot{G}|/\Sigma[X_i]$ and $\Delta\mu\Delta\dot{\mu}/RT$ calculated for the Michaelis–Menten reaction model, to demonstrate Eq. (43).

**Figure 7:** (Color online) Time evolution of $|\dot{G}|/n_{sol}$ and $\Delta\mu\Delta\dot{\mu}/RT$ calculated for the Belousov–Zhabotinsky reaction model, to demonstrate Eq. (52).

**Figure 8:** (Color online) Time evolution of $|\dot{G}|/n_{sol}$ and $\Delta\mu\Delta\dot{\mu}/RT$ calculated for the Michaelis–Menten reaction model, to demonstrate Eq. (52).

**Table 1:** Summary of the bases of the chemical potential and standard deviation in the thermodynamic uncertainty relations presented in the present study, Refs. 17 and 18.

**Figures and Tables**

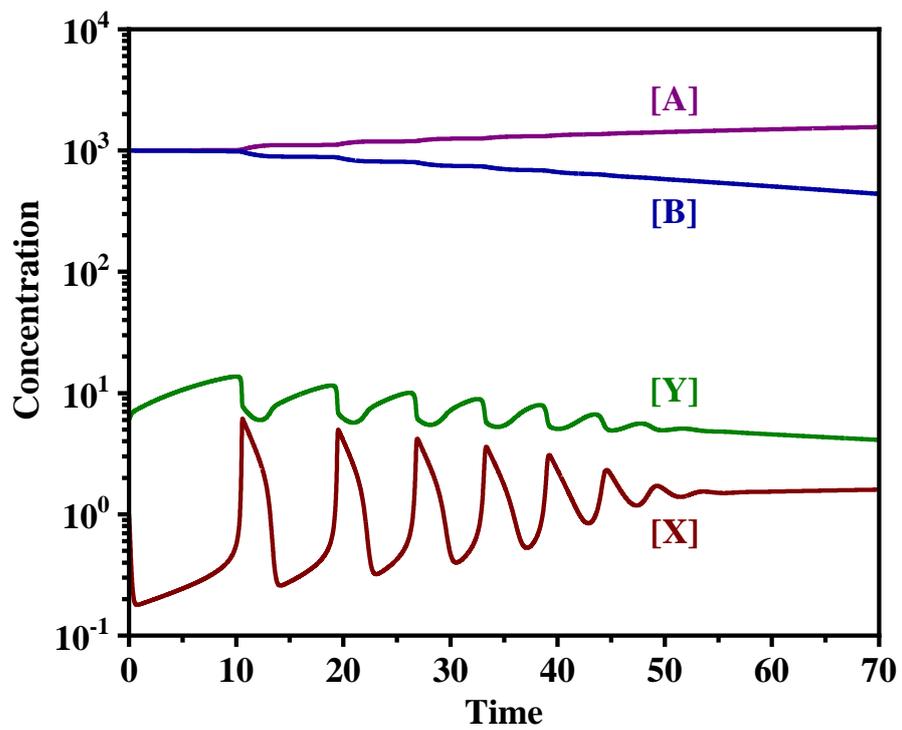

Fig. 1

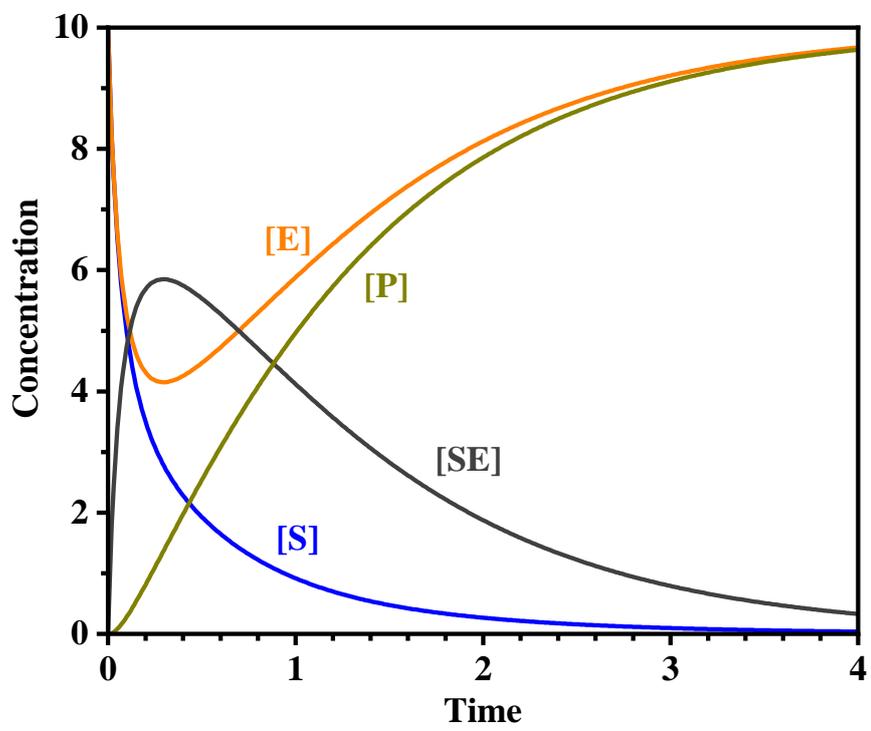

Fig. 2

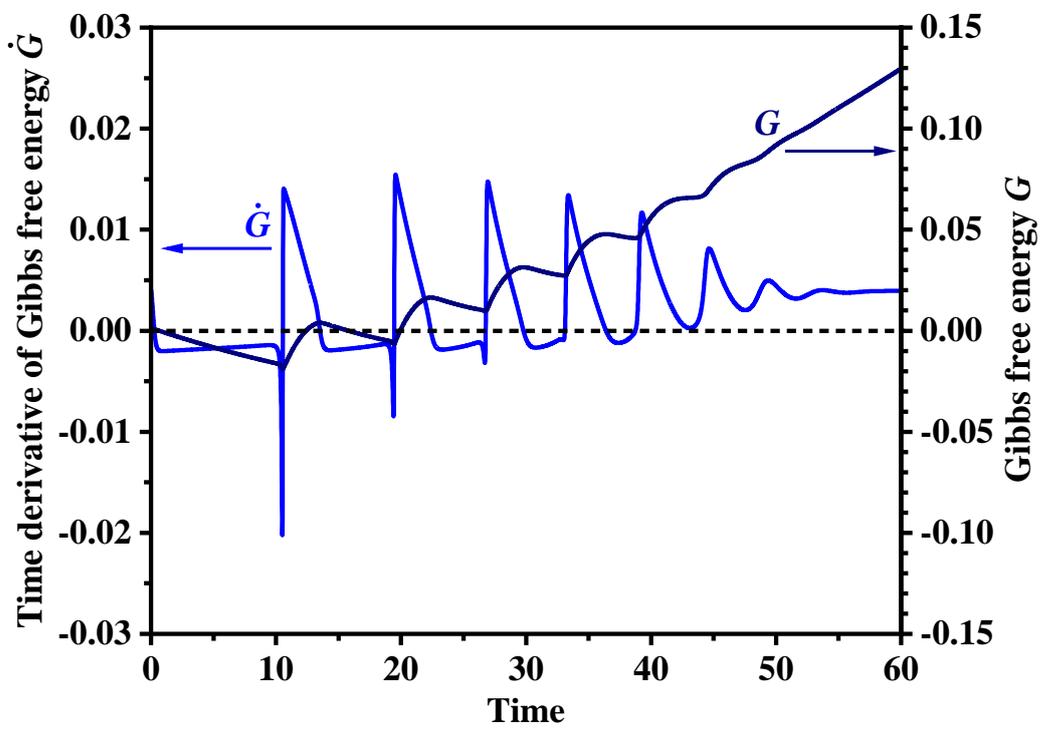

Fig. 3

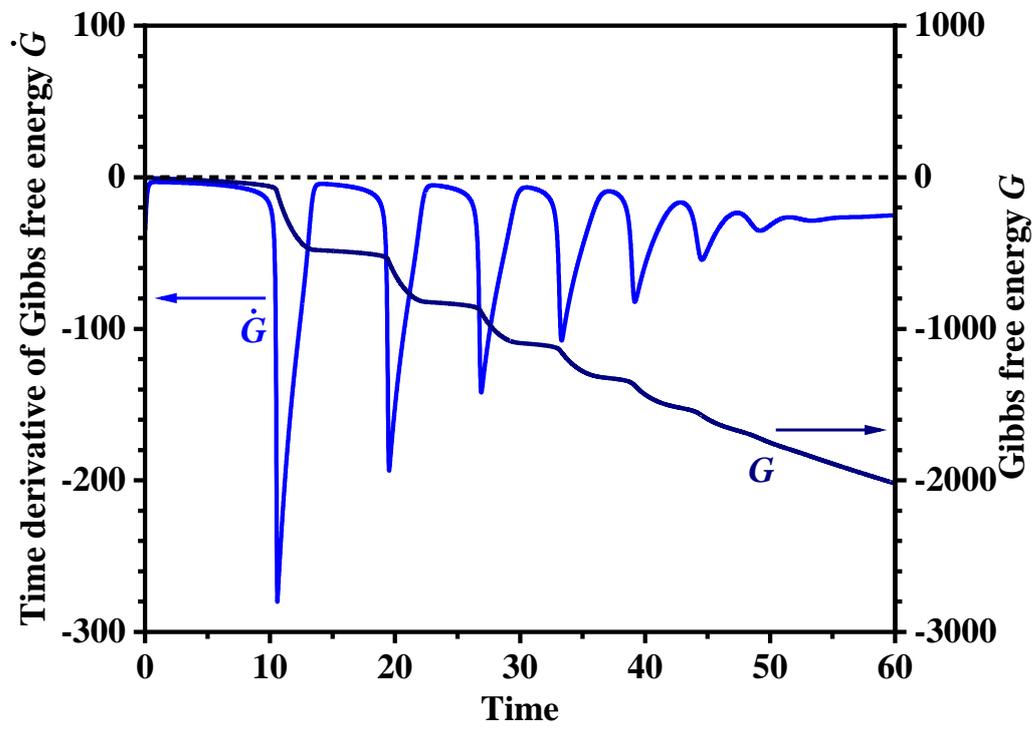

Fig. 4

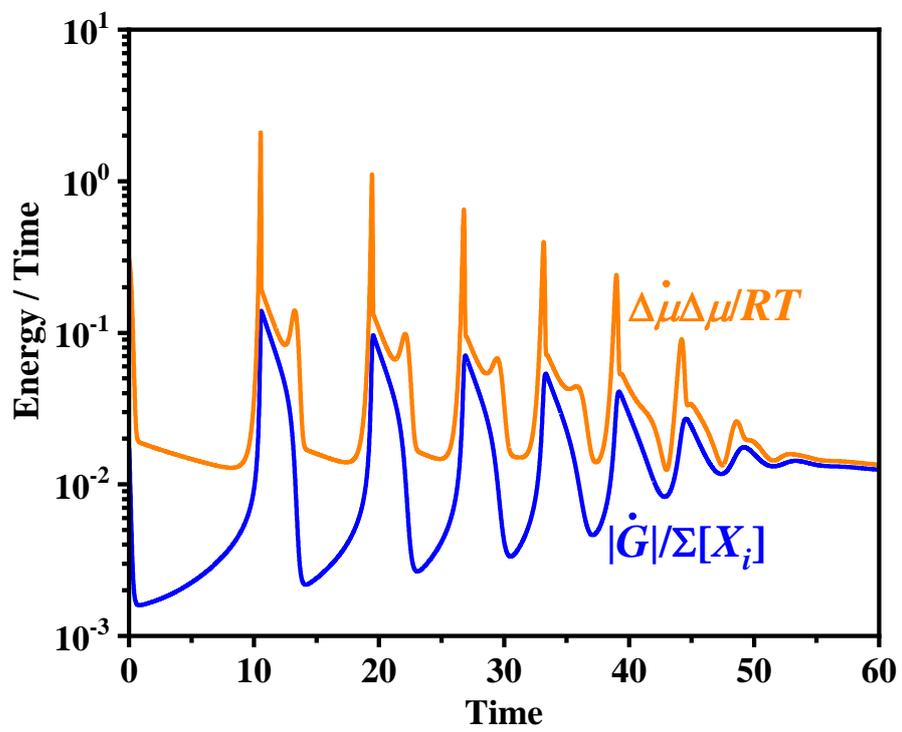

Fig. 5

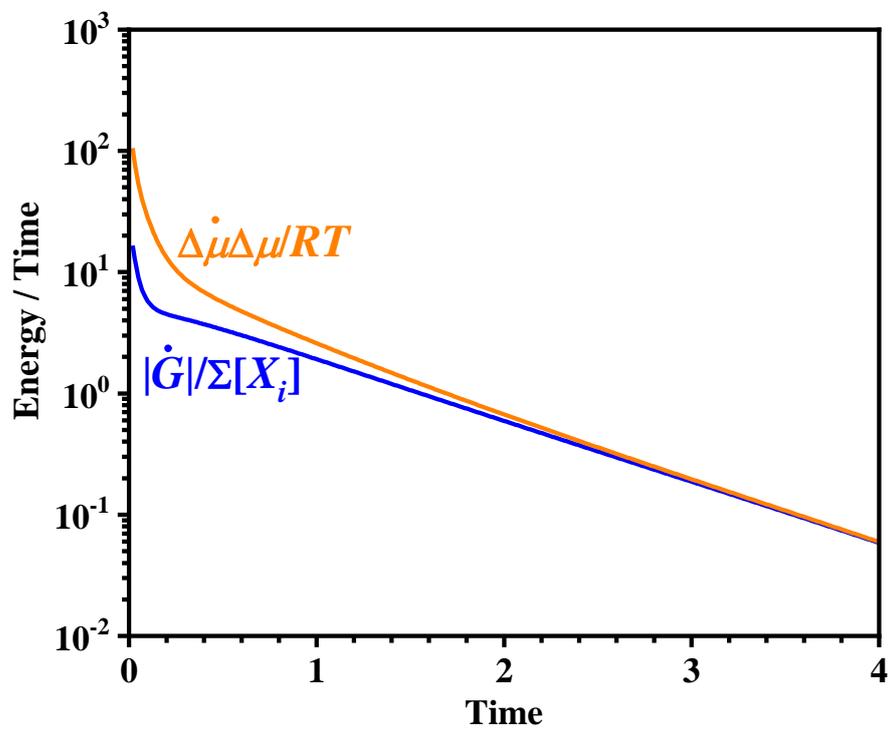

Fig. 6

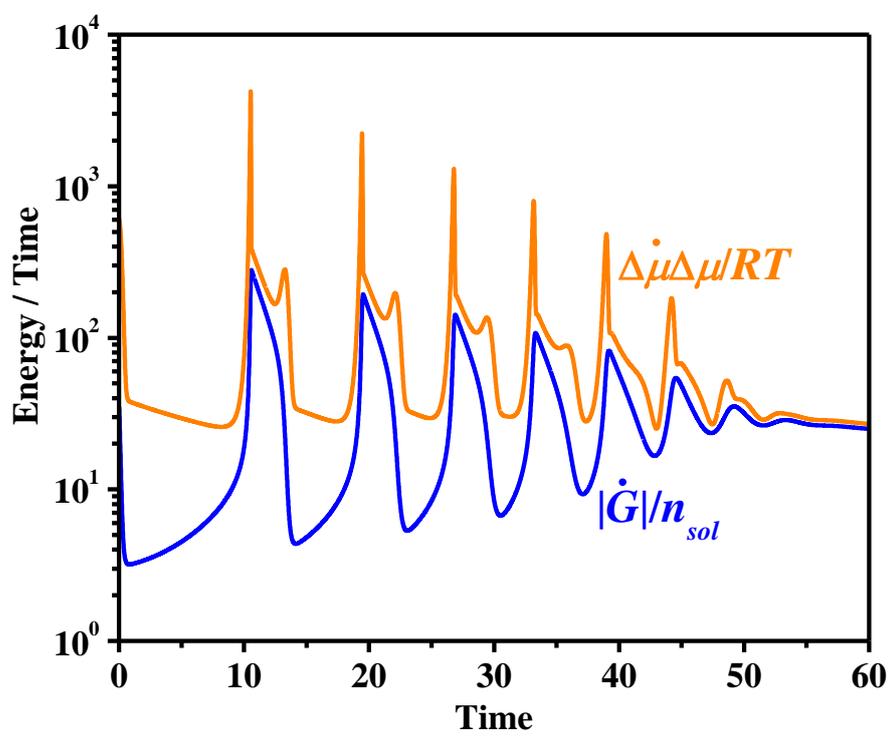

Fig. 7

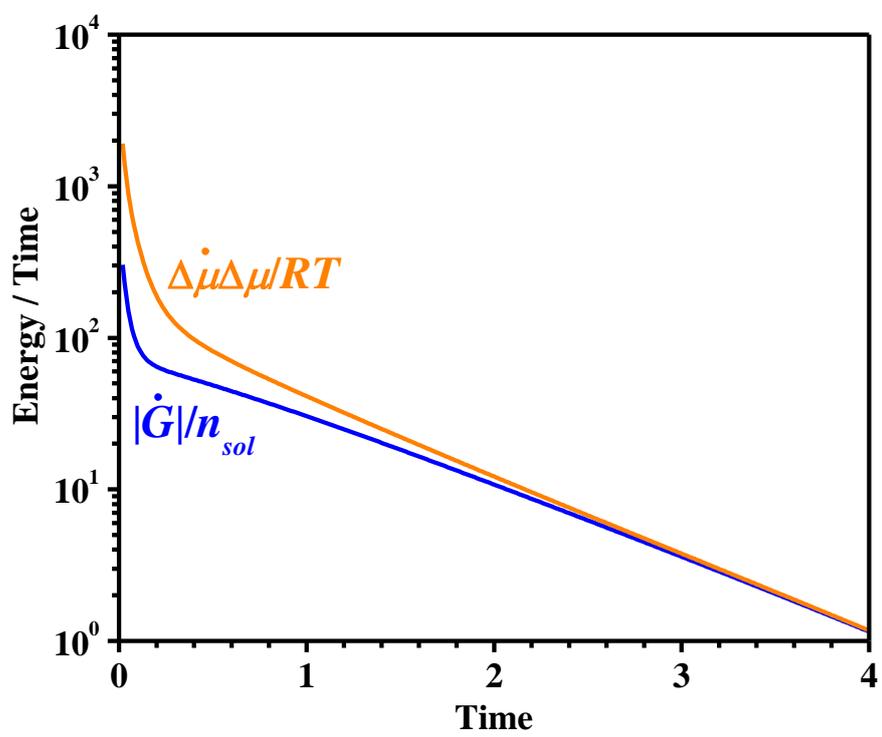

Fig. 8

| Thermodynamic uncertainty relation | Chemical potential | Standard deviation |
| --- | --- | --- |
| Eq. (43), this work | Concentration | Molar fraction |
| Eq. (52), this work | Molar fraction | Molar fraction |
| Eq. (58), this work | Concentration | Concentration |
| Ref. 17 | Concentration | Concentration |
| Ref. 18 | Concentration | Molar fraction |

Table 1